\documentclass[12pt,preprint]{aastex}
\usepackage{fullpage,amsmath,amssymb,latexsym}
\usepackage{amsmath,amssymb,latexsym}
\usepackage{url}
\shorttitle{The Uncertainty in Saturn's Shape and Rotation Period}
\shortauthors{Helled \& Guillot}

\begin{document}
\title{Interior Models of Saturn:\\
Including the Uncertainties in Shape and Rotation }
\author{Ravit Helled$^1$ and Tristan Guillot$^2$}
\affil{$^1$Department of Geophysics, Atmospheric and Planetary Sciences\\
Tel-Aviv University, Tel-Aviv, Israel\\
$^2$Universit\'e de Nice-Sophia Antipolis, Observatoire de la $C\hat{o}te$ $d'Azur$, CNRS UMR 7293, {\it BP 4229} 
{\it06304 Nice, France}}

\begin{abstract}
The accurate determination of Saturn's gravitational coefficients by Cassini could provide tighter constrains on Saturn's internal structure. Also, occultation measurements provide important information on the planetary shape which is often not considered in structure models. 
In this paper we explore how wind velocities and internal rotation affect the planetary shape and the constraints on Saturn's interior.  
We show that within the geodetic approach (Lindal et al., 1985, ApJ, 90, 1136) the derived physical shape is insensitive to the assumed deep rotation. Saturn's re-derived equatorial and polar radii at 100 mbar are found to be 54,445 $\pm$10 km and 60,365$\pm$10 km, respectively. 
To determine Saturn's interior we use {\it 1 D} three-layer hydrostatic structure models, and present two approaches to include the constraints on the shape. These approaches, however, result in only small differences  in Saturn's derived composition. 
The uncertainty in Saturn's rotation period is more significant: with Voyager's 10h39mns period, the derived mass of heavy elements in the envelope is 0-7 M$_{\oplus}$. With a rotation period of 10h32mns, this value becomes $<4$ $M_{\oplus}$, below the minimum mass inferred from spectroscopic measurements. 
Saturn's core mass is found to depend strongly on the pressure at which helium phase separation occurs, and is estimated to be 5-20 M$_{\oplus}$. Lower core masses are possible if the separation occurs deeper than 4 Mbars. 
We suggest that the analysis of Cassini's radio occultation measurements is crucial to test shape models and could lead to constraints on Saturn's rotation profile and departures from hydrostatic equilibrium.

\end{abstract}

 
\section{Introduction}
Knowledge of giant planets' internal structures (composition, the material distribution with radial distance, and core mass) is crucial for understanding giant planet origin.  
The internal structures of the gas giant planets in our solar system, Jupiter and Saturn, are constrained by measurements of their physical parameters such as mass, radius, rotation period, and gravitational moments. 
Although Cassini's accurate measurements of Saturn's gravitational field offer an opportunity to better constrain Saturn's internal structure, the realization that Saturn's rotation period is not well constrained within a few minutes introduces an uncertainty that must be considered when modeling the planetary interior.  \par
The interior of a giant planet is typically derived for a given solid-body rotation period (Zharkov and Trubitsyn, 1978; Guillot, 2005) and the inferred model (i.e., composition and its depth dependence) is therefore dependent on the assumed rotation period and on the assumption of solid-body rotation. The uncertainty in the planetary rotation also introduces uncertainty in the planetary hydrostatic shape which enters the interior models via the mean/equatorial radius (Helled et al., 2008, 2011). So far interior models of Saturn assumed equatorial radius and solid-body rotation rate which result in an incorrect mean (volumetric) radius. In this paper we suggest how the uncertainty in Saturn's rotation period/state and its corresponding shape should be treated when deriving structure models. \par

\section{Saturn's Rotation Period}

It has been acknowledged that Saturn's internal rotation period is unknown within a few minutes since Cassini's SKR measured a rotation period of 10h 47mns 6s (Gurnett et al., 2007), longer by about eight minutes than the Voyager radio period of 10h 39mns 22.4s (Ingersoll and Pollard, 1982; Dessler, 1983). In addition, during Cassini's orbit this period was found to be changing with time (Gurnett et al., 2007, 2009). As a result, it is clear that the SKR measurements do not represent the rotation period of Saturn's deep interior. Due to the alignment of the magnetic pole with the rotation axis, Saturn's rotation period cannot be obtained from magnetic field measurements (Sterenborg and Bloxham, 2010). Several groups have attempted to determine the rotation period of Saturn's deep interior using various methods.  Anderson and Schubert (2007 -hereafter AS07) have argued that Saturn's  rotation period can be found by minimizing the dynamical heights of its isobaric surface, and suggested a rotation period of 10h 32mns 35s. Read et al. (2009) have analyzed the latitudinal distribution of potential vorticity on Saturn and derived a rotation period of 10h 33mns 13s based on dynamical arguments. A shorter rotation period for Saturn results in  atmospheric dynamics which resembles Jupiter's with smaller wind velocities. \par
 
 \section{The Planetary Shape}
 The shape of a rotating planet in hydrostatic equilibrium, also known as the {\it reference geoid}, is defined as the level surface of equal effective potential $U$ by
\begin{equation}
U = \frac{G M}{r} \left( 1 - \sum_{n=1}^\infty \left( \frac{a}{r} \right)^{2 n} J_{2 n} P_{2 n} \left( \cos \theta  \right)  \right)+ \frac{1}{2} \omega^2 r^2 \sin^2 \theta.
\label{U}
\end{equation}
The level surface can be computed up to different orders in the smallness parameter $q=\omega^2 a^3/GM$, where $a$ is the geoid's equatorial radius and $GM$ is its mass multiplied by the gravitational constant. The radius $r$ as a function of cos$\theta$ can be expanded by $
r({\cos}\theta)= a\left( 1+ \sum_{n=1}^{\infty}r_n({\cos}\theta) q^n\right)$. $\theta$ is the co-latitude and $\omega$ is the angular velocity given by $2\pi/P$, with $P$ being the (solid-body) rotation period. Saturn's harmonic coefficients $J_{2n}$ as computed by Jacobson et al. (2006) for a reference equatorial radius of 60,330 km in units of $\times 10^6$ are $J_2=16,290.71\pm0.27$, $J_4=-935.8\pm2.8$, and $J_6=86.1\pm9.6$.  
\par

The measured shape of the planet, however, is also affected by atmospheric winds.  The contribution of the winds to the shape, also known as the 'dynamical heights' (Lindal et al., 1985) can be added to the reference geoid shape in order to reproduce the physical shape (reference geoid + dynamical heights) of the planet.  By using zonal wind data with respect to an assumed solid-body rotation period, the dynamical heights with respect to the reference geoid are given by (Lindal et al., 1985)
 \begin{equation}
h(\phi) = \frac{1}{g}\int_{\phi}^{\frac{\pi}{2}}V_W(\phi)\left(2\omega_{\mathrm{ref}}+\frac{V_W}{r_{\mathrm{ref}}(\phi)\cos\phi}\right)\frac{\sin(\phi+\psi_{\mathrm{ref}})}{\cos\psi_{\mathrm{ref}}}r_{\mathrm{ref}}(\phi)d\phi
\end{equation}
where $g$ is the average gravity acceleration, $\psi$ is a small angle that gives the difference between the planetocentric latitude $\phi$ and the planetographic latitude $\phi'$, $\psi =\phi'-\phi$, and $V_W(\phi)$ is the zonal wind velocity. Parameters with the subscript 'ref' represent quantities calculated for the reference geoid. \par

There are two important assumptions behind these two equations. First,  writing eq.(1) implicitly implies that the centrifugal acceleration derives from a potential, which is possible only for solid-body rotation ($\omega=$ constant) or differential rotation on cylinders ($\omega=\omega(r\cos[\theta]$)). As a consequence, the wind speed at the poles must be zero: $V_W(\phi=\pi/2)=0$. Second, it assumes that the gravitational coefficients $J_{2n}$ defined by eq.(1) are set to their measured value, independently of the considered pressure level. \par

It is important to note that all interior models of the giant planets published thus far are {\it hydrostatic} and {\it one-dimensional} (1D). There are only two classes of interior models that are hydrostatic. The first and by far most common assumes solid-body rotation (e.g., Guillot, 2005) and the second differential rotation on cylinders all the way to the center (Zharkov and Trubitsyn, 1978; Hubbard 1999). As a result, there is a whole class of interior models that are not considered, including interior models which combine solid-body rotation in the deep interior and an outer shell which rotates differentially on cylinders as often assumed in dynamical models (Heimpel and Aurnou, 2007; Kaspi et al., 2010). Such models are necessarily non-barotropic and cannot be calculated using potential theory. \par

The physical parameters that are used to constrain the planetary interior are the planet's total mass, its rotation period, the measured gravitational harmonics and the planet's equatorial radius $R_{eq}$ or mean radius $\overline{R}$ (see Guillot, 2005, Fortney and Nettelmann, 2010 and references therein). As discussed earlier, while today more accurate measurements of Saturn's gravitational field from the Cassini spacecraft are available, the realization that Saturn's rotation period is not well constrained and the fact that interior models do not account for the contributions from the atmospheric winds and shape data introduce an additional uncertainty to interior models. 
Although Saturn's shape is determined independently of rotation period by occultations, the measured radii include the contributions from the wind and therefore should {\it not} be compared with the equilibrium shape of a solid-body rotating planet as derived by interior models. The equatorial radius of Saturn is affected by the large wind velocities near the equator. An alternative could be fitting the interior models to the polar radius, since Saturn's polar radius is expected to be fairly independent of rotation period and atmospheric winds, however, there are no available occultation measurements of Saturn's poles. In addition, Saturn's north-south asymmetry introduces another uncertainty in determining its polar radius (Lindal et al., 1985). As a result, interior models should account for the uncertainty in the polar, mean, and equatorial radii of Saturn, and should not use the measured values which correspond to the physical shape of the planet.\par

We next investigate whether the formulation of the physical shape (geodetic approach) is independent of rotation period. We compute Saturn's physical shape for three different solid-body rotation periods: Voyager radio period of 10h 39mns 22.4, 10h 45mns 24s and AS07 period of 10h 32mns 35s. The wind velocities are taken from Sanchez-Lavega et al. (2000). The dynamical heights as a function of latitude for the three rotation periods are shown in the upper panel of Figure 1. The red, blue and black solid lines represent Voyager's period, 10h 45mns 24s, and AS07 period, respectively. Since the wind velocity data do not go all the way to the poles, at high latitudes the wind velocities are set to zero. The dark color circles represent radii obtained by radio occultation measurements with the Pioneer 11, Voyager 1, and Voyager 2 spacecrafts (Lindal et al., 1985) assuming an error bar of 7 km. The circles with the light colors are stellar occultation measurements as derived by Hubbard et al. (1997). These measurements are found to be in excellent agreement with the radio occultation measurements. Due to the agreement between the two data sets, and since the stellar occultation measurements correspond to a relative narrow region, and need to be adjusted to correspond to the 1 bar pressure-level, a process which adds additional uncertainty, hereafter, our analysis uses only the radio occultation measurements. Since the two agree well, we don't expect a change in the results when stellar occultation measurements are also considered. The dynamical heights are larger for longer rotation periods, and for AS07 rotation period they are minimized. For simplicity, hereafter we set Saturn's continuous physical shape (reference geoid + dynamical heights) to the shape of {\it a reference geoid with AS07 rotation period}. The radius residuals (derived physical shape compared to AS07 geoid) vs. latitude are shown in the bottom panel of Figure 1.  As can be seen from figure, the derived physical shapes of Saturn for rotation periods of 10h 39mns 22.4 and 10h 45mns 24s are roughly the same, but not identical. These small differences are due to fact that the adopted rotation profiles differ at the poles where we have no data. We can therefore conclude that for an uncertainty of several minutes in rotation period the physical shape as derived from the geodetic approach is essentially independent of rotation period. In that regard, it is important to note that the formulation of Lindal et al. (1985) for the physical shape is not identical to the computed shape when differential rotation on cylinders all the way to the center is assumed. \par

We also derive the radius residuals for the shape of a geoid with the Voyager rotation period and the equatorial radius reported by Lindal et al. (1985) as typically assumed by previous interior models. The residuals are found to be small near the equator but increase up to 90 km at the polar regions. The $\sim$100 km difference in Saturn's polar radius is not surprising given that the dynamical heights of Saturn with the Voyager radio period are of the same order. We therefore suggest that this combination, which was often assumed by interior modelers, is {\it inconsistent with the available data}. Finally, we suggest that Saturn's equatorial, polar and mean radii at the 100 mbar pressure-level are 60,365$\pm$10 km,  54,445 $\pm$10 km and 58,323$\pm$10 km, respectively. These values are in good agreement with the values derived by Lindal et al. (1985). 
\par

\section{Saturn's Interior Models}

\subsection{Accounting For The Planet's Shape} 

The uncertainties in Saturn's internal rotation and shape should be included in structure models as this can affect the resulting gravitational potential, and thus what we can infer from its measurement (Hubbard, 1982; Guillot et al., 1999; Hubbard et al., 1999; Militzer et al., 2008; Kaspi et al., 2010).  Until 2D/3D interior models become available, we propose to account for the uncertainties in the rotation state and shape by increasing the formal uncertainties as we discuss below. In essence, this approach is similar to what had been proposed by Hubbard (1982) and then used by Guillot and collaborators (Guillot et al., 1997, Guillot 1999) to constrain the structures of Jupiter and Saturn, but now we also consider the non-negligible uncertainties that arise from the determination of the planetary shape.  \par

Up to now, all interior models of Saturn have been calculated using Saturn's measured equatorial radius and a solid-body rotation set to Voyager's system III value. As we have seen in the previous section, this leads to Saturn models with a polar radius that is off by about 120 km compared to the observations, and hence, to a wrong mean radius. While the effect is small, it is much larger than the observational uncertainties so that it is important to quantify it. 

Because we use interior models that assume solid-body rotation while the planet's shape is affected directly by differential rotation (the planet's latitudinal wind profile), there is necessarily a degree of arbitrariness in the solutions that we search for. On one hand, we could consider that most of the planetary mass rotates as solid-body so that the gravitational field is essentially set by the corresponding geoid while the planet's shape is defined by a shell of comparably much smaller mass in which differential rotation is dominated. On the other hand, we could assume that the observed winds penetrate deep into the planetary interior and involve a large fraction of the mass. In that case, it would seem natural to use the mean radius as the conserved quantity when using models that assume solid-body rotation. 

These different situations are depicted in Figure 2 and represent the different cases considered in this work. Case (0) corresponds to previous studies in which Saturn's equatorial radius was held equal to the observed value. Case (1) corresponds to the opposite situation in which the polar radius is held equal to its measured value (54,438 km at the 100 mbar pressure-level -- Lindal et al., 1985). This case appears more realistic because the observed equatorial bulge can then be attributed to the observed winds, confined to a limited outer shell. Finally, Case (2) corresponds to fixing the mean radius to its observed value. 
These cases might not represent realistic configurations but they are useful in defining the range of Saturn's equatorial/mean radius for an interior modeler. It is desirable that future models will calculate the shapes self-consistently when various rotation profiles are assumed. \par

The AS07 rotation period was derived by a method that minimizes the dynamical heights which in our case essentially minimizes the differences between Case (1) and Case (2), and therefore for this rotation period the two cases converge. Saturn's physical shape at the 100 mbar pressure-level is shown in Figure 3. The black curve corresponds to a reference geoid with AS rotation period, with the black dots being the published radio occultation measurements. The red curves correspond to reference geoid shapes with Voyager's radio period; the dashed and solid red lines correspond to Case (1) and Case (2), respectively. Saturn's measured radii fit well a reference geoid with the AS07 rotation period due to the minimization of the dynamical heights. While a reference geoid with the Voyager radio period results in large residuals in the equatorial region when Case (1) is considered, a very good agreement is found for Case (2). 
\par

Saturn's equatorial (top) and polar (bottom) radii as a function of an assumed solid-body rotation period for the two cases are shown in Figure 4. The blue, black and red curves correspond to Case (0), Case (1) and Case (2), respectively. Case (1) and Case (2) meet at the AS07 rotation period. Clearly, Case (0) overestimates the polar radius due to the fixed value of the equatorial radius to its measured value, we therefore argue that Case (0) should no longer be assumed. Case (2) results in a more moderate change in the equatorial radius copmared to Case (1), since is that case both the polar and equatorial radii are modified to fit the measured planetary volume. In Case (1) the polar radius is held fix which requires a more significant modification of the equatorial radius in order to fit Saturn's measured volume. We suggest that the uncertainty in Saturn's equatorial (polar) radius for a given solid-body rotation period should be taken as the (absolute) {\it difference} between the equatorial (polar) radius in Case (1) and Case (2). Since the dynamical heights are up to the order of 100 km, the uncertainty in Saturn's equatorial/polar radius can be of the same order. 
\par

\subsection{Model Hypotheses}
In this section we investigate how the uncertainty in Saturn's shape and rotation period/state affects its derived internal structure and bulk composition. 
To model the planetary interior we use a standard one-dimensional three-layer hydrostatic interior model in which the planet is assumed to consist of a central ice/rock core and an envelope which is split into a helium-rich metallic hydrogen region and a helium-poor molecular region that is connected to the atmosphere. The interior models are computed assuming the following: The temperature at the 1-bar pressure-level is assumed to range between 130 and 145 K. The helium mass fraction at the outer envelope ranges between 0.11 and 0.25, with the overall helium mass fraction of 0.265-0.275. The pressure in which the transition from the helium-rich to helium-poor occurs ($P_{{\text{transition}}}$) is assumed to be between 1 and 4 Mbars, corresponding roughly to the range of values expected from calculation of mixtures under high pressures (Morales et al. 2009). The heavy elements besides the discontinuity in the helium abundance at $P_{{\text{transition}}}$ are assumed to be homogeneously mixed within the planetary envelope. 

Finally, in order to account for the additional uncertainty linked to differential rotation, we follow Hubbard (1982) and set the uncertainty in the gravitational harmonics to be $\delta J_2$ = -80$\times10^{-6}$, $\delta J_4$ = +20$\times10^{-6}$, $\delta J_6$ = +10$\times10^{-6}$. We note that Kong et al. (2012) have recently investigated the corrections to the gravitational coefficients due to both zonal winds and oblateness in a self-consistent manner using a general perturbation theory in which oblate spherical coordinates are used. Future work should include the fact that the corrections in the various gravitational moments are correlated, with the possibility to narrow down the ensemble of possible solutions. 
 
We use two sets of gravitational data, the first, hereafter Voyager's $J$s, are taken to be $J_2 = 16258\pm41\times10^{-6}$, $J_4 = -905\pm41\times10^{-6}$, and $J_6 = 98\pm51\times10^{-6}$. Cassini's $J$s are set to be $J_2 = 16251\pm40 \times10^{-6}$, $J_4 = -926\pm11\times10^{-6}$, and $J_6 = 81\pm11\times10^{-6}$.  The ``1 sigma'' uncertainties are calculated as the geometric mean of the measurement uncertainty and of the difference between the results assuming solid-body and rotation-on-cylinders. The model gravitational moments are calculated for the 60,330 km reference radius for comparison with the observations. More details on the interior model can be found in Guillot (2005) and references therein. \par

We consider two solid-body rotation periods: the Voyager radio period (labelled ``slow''), and the AS07 rotation period (labelled ``fast''). For the Voyager period we consider the three cases described previously. The corresponding model 1-bar equatorial radius is 60,269 km for Case (0), 60,148 km for Case (1) and 60,238 km for Case (2). For each case, the uncertainty is assumed equal to 10\,km. For the AS07 rotation period, the three cases lead to similar values and we therefore only consider a model in which the 1-bar equatorial radius is $60\,238\pm 10$\,km.

\subsection{Consequences Of The Shape Models and Improved Gravitational Fields}

First, we investigate how the different assumptions on the planetary shape and the $J$s measurement uncertainty affect the derived internal structure. Figure 5 presents a comparison of the derived interior model solutions with $P_{{\text{transition}}}= 1$\,Mbar and for the Voyager rotation period, with our different assumptions. The contours correspond to interior models that fit within 2 sigmas in equatorial radius, $J_2$, and $J_4$. In all cases, $J_6$ was found to fit within 1 sigma of the observed value. 

The most significant change results in from the improved determination of Saturn's gravitational field obtained from Cassini over the previous ``Voyager'' values (obtained from a combination of the Pioneer and Voyager spacecraft measurements). As seen from Figure 5, the new gravitational moments exclude core masses over 20\,M$_\oplus$ (instead of 25\,M$_\oplus$ previously), and also lead to generally smaller masses of the heavy elements in Saturn's envelope (from less than 6.5\,M$_\oplus$ to less than 5.25\,M$_\oplus$). 

In comparison, the effect of the planetary shape is rather limited: its effect on the inferred core masses is minor, and only leads to a $\sim 1\,\rm M_\oplus$ added uncertainty in the inferred mass of heavy elements in the envelope. The resulting gravitational moments are barely changed. This could be understood by the fact that any change in the planet's mean radius may be offset by an adjustment of the composition of the envelope with little consequence for the planet's gravitational field.

\subsection{Model Results With Different Rotation Periods}

We next explore the effect of the assumed solid-body rotation period on the inferred composition. Figure 6 shows the interior model solutions when using Cassini's $J$s for the Voyager rotation period (``slow'') and for the AS07 one (``fast''). The solutions are provided as a function of the pressure at which the transition from a helium-poor to a helium-rich envelope occurs, between 1 and 4\,Mbar. 

For the standard Voyager (``slow'') case, a large variety of solutions is found. The core mass and amount of heavy elements in the envelope are both found to depend crucially on the helium transition level. For low values (around 1Mbar), a relatively large core mass (between 10 and 20\,M$_\oplus$) and a small amount of heavy elements in the envelope (less than about 5\,M$_\oplus$) are required. For larger values of $P_{{\text {transition}}}$, the core mass decreases while the heavy element masses in the envelope increases. Once the transition pressure reaches a value of $P_{{\text{transition}}}=4\,$Mbar, only a small set of solutions is found. At this point, the jump in helium abundance required to fulfill the protosolar helium to hydrogen constraint becomes large and the helium-rich envelope has an effect on the planet's $J_2$ that is very similar to that of a rock/ice core. This thus explains the smaller core value, as proposed by Fortney and Hubbard (2003). Although solutions with core masses smaller than 4\,M$_\oplus$ were not found, we believe that this is due to the numerical algorithm used and that {\it solutions with no core are possible}, as in the case of Jupiter (Guillot et al. 1997, 1999; Kramm et al., 2011). However, small/no-core solutions generally corresponds to a relatively deep transition compared to the results of numerical experiments (Morales et al. 2009). 

The ensemble of valid solutions is found to be much more limited in the ``fast'' AS07 rotation period case. First, solutions are found only in the case of a transition pressure at 1 or 2\,Mbar. Second, the solutions span a range of heavy element masses in the envelope that is typically about half of the values found using the Voyager rotation period. The choice of the deep rotation period used for the interior models thus has a significant effect on the inferred composition of the planet. 

It is instructive to compare the mass of heavy elements obtained in the envelope with the known atmospheric composition (Guillot \& Gautier, 2006 and references therein), especially given our assumption of a uniform abundance of heavy elements in Saturn's envelope. The grey areas in Figures 5 and 6 represent a ``forbidden zone'' with abundances that are lower than the spectroscopic determination. All model solutions for the AS07 rotation period are in that zone and are hence excluded. Solutions with the AS07 rotation period are likely to be found when a discontinuity in the heavy elements distribution in the envelope is allowed (see Fortney and Nettelmann 2010), due to the additional degree of freedom that is then introduced. A discontinuity in the heavy elements distribution could be a result of double diffusive convection (e.g., Leconte and Chabrier, 2012) but at this point it is not clear what assumption on the heavy element distribution is more appropriate when modeling Saturn's interior. 
Interestingly, the abundance of water appears to be limited to 8 times the solar value, i.e. at most comparable to the derived enrichment in carbon. 

Two caveats are important however: first we did only consider one equation of state for the hydrogen-helium mixture (Saumon et al., 1995). Although the effect of different H-He EOSs appear limited in the case of Saturn (see Saumon \& Guillot 2004), it is not completely negligible. Future work should consider models calculated with recent EOSs. Second but more importantly, our rough treatment of differential rotation means that a fine analysis of the results should be deferred to 2D or 3D interior models.

\subsubsection*{Jupiter}
For Jupiter, interior models that use its physical measured shape are more consistent. This is because Jupiter's dynamical heights are fairly small and are of the order of 5 km (Ingersoll, 1970; Helled et al. 2009). The shape of Jupiter's reference geoid with a rotation period of 9h 55m 29.7s (System III) fits well to the occultation measurements. We suggest that assuming an uncertainty of about 10 km in Jupiter's radius when computing static interior models is sufficient.
The main uncertainties in determining the bulk composition of Jupiter remain the depth of differential rotation, the global water abundance, and the distribution of the heavy elements within its interior. Constraints from the upcoming {\it Juno} mission are expected to narrow the parameter-space of possible internal structures. 

\section{Discussion and Conclusions}

While accurate measurements of Saturn's gravitational field can provide tighter constrains on Saturn's interior structure, the uncertainty in Saturn's rotation rate, and more importantly, rotation state lead to uncertainty in its bulk composition. In addition, we suggest that occultation measurements can provide valuable information on the planetary shape which is often not considered when deriving interior models.  
We used the geodetic approach as presented by Lindal et al. (1985) and showed that the derived physical shape of Saturn using this approach is independent of rotation period. We also showed that using the combination of Voyager's rotation period with physical shape data, as typically assumed by interior modelers, is inconsistent due to the contribution of the winds to the measured shape. We re-derived Saturn's shape at the 100 mbar pressure-level and found that Saturn's equatorial, polar and mean radii are 60,365$\pm$10 km,  54,445 $\pm$10 km and 58,323$\pm$10 km, respectively.  \par 

We next explored how the uncertainties in internal rotation and shape affect the derived internal structure of Saturn by using a standard 1D three-layer interior model.  We found that when the ``slow'' Voyager period of 10h39mns is assumed for the deep interior, Saturn's heavy element mass in the envelope is between 0 and 7 M$_{\oplus}$, while the core mass ranges between 5 and 20 M$_{\oplus}$. The inferred core mass was found to decrease significantly when the pressure in which helium separation occurs is higher. Knowledge on the pressure in which  the transition between molecular and metallic hydrogen occurs, is therefore crucial, for better constraining Saturn's interior. We suggest that interior models of Saturn with no ice/rock core are possible. Finally, we showed that under our model assumptions, interior models with a ``fast'' rotation period of 10h32mns have heavy element mass in the envelope which is too small compared with atmospheric spectroscopic measurements. Solutions with the ``fast'' rotation period, however, are expected to be found when a discontinuity in the heavy elements distribution in the envelope is permitted, although there is no clear physical reason to favor such a possibility. \par 

The lack of knowledge on the depth of differential rotation in Saturn is a major source of uncertainty on the internal structure and global composition of the planet. We suggest that an analysis of Cassini's radio occultation measurements is crucial to test the shape models and could lead to constraints on Saturn's rotation profile and departures from hydrostatic equilibrium. This could usefully complement the measurement of high-order as well as odd gravitational moments that would constrain the depth of differential rotation as expected from the {\it Juno} and {\it Cassini Solstice} missions (Hubbard 1999, Kaspi 2013).

Finally, this study suggests that in order to make full use of {\it Cassini} and {\it Juno} data, non-hydrostatic contributions and various rotation profiles should be incorporated self-consistently in new interior structure models that do not require the centrifugal acceleration to derive from a potential. \par

\subsection*{Acknowledgments} 
We thank Yohai Kaspi for discussions and sharing results in advance of publication. We also wish to thank Bill Hubbard for his valuable comments. TG acknowledges support from {\it CNES} and the {\it Programme National de Planétologie}.


\section*{References}

Anderson, J. D. and Schubert, G. 2007, SaturnÕs Gravitational Field, Internal Rotation, and Interior Structure. Science, 317, 1384--1387.

Asplund, M., Grevesse, N., Sauval, A. J. and Scott, P., 2009. The Chemical Composition of the Sun. Annual Review of Astronomy \& Astrophysics, 47, 481.

Dessler, A. J., 1983. Physics of the Jovian magnetosphere. Cambridge University Press, New York, 498.

Fortney, J. J. and Hubbard, W. B., 2003. Phase separation in giant planets: inhomogeneous evolution of Saturn. Icarus, 164, 228. 

Fortney, J. J. and Nettelmann, N., 2010. The Interior Structure, Composition, and Evolution of Giant Planets. Space Science Reviews, 152, 423. 

Guillot, T., Gautier, D., Hubbard, W. B., 1997. NOTE: New Constraints on the Composition of Jupiter from Galileo Measurements and Interior Models. Icarus, 130, 534.

Guillot, T., 1999. Interiors of giant planets inside and outside the Solar System. Science, 296, 72.

Guillot, T., 2005. The Interiors of Giant Planets: Models and Outstanding Questions. Annual Review of Earth and Planetary Sciences, 33, 493.

Guillot, T. and Gautier D., 2007. Giant Planets. In: {\it Treatise on Geophysics}. Planets and Moons. Schubert G., Spohn T. (Ed.) (2007) 439. 

Gurnett, D. A., Persoon, A. M., Kurth, W. S., Groene, J. B., Averkamp, T. F., Dougherty, M. K. and  Southwood, D. J. 2007, The Variable Rotation Period of the Inner Region of SaturnÕs Plasma Disk.  Science, 316, 442.

Gurnett, D. A., Persoon, A. M., Groene, J. B., Kopf, A. J., Hospodarsky, G. B., Kurth, W. S., 2009. A north-south difference in the rotation rate of auroral hiss at Saturn: Comparison to Saturn's kilometric radio emission. GRL, 36, L21108.

Heimpel, M. and Aurnou, J., 2007. Turbulent convection in rapidly rotating spherical shells: A model for equatorial and high latitude jets on Jupiter and Saturn. Icarus, 187, 540.

Helled R., G. Schubert, and J. D. Anderson, 2009. Jupiter and Saturn rotation periods. Planet. Space Sci., 57, 1467. 

Hubbard, W. B., 1982. Effects of Differential Rotation on the Gravitational Figures of Jupiter and Saturn. Icarus, 52, 509. 

Hubbard, W. B., Porco, C. C., Hunten, D. M., Rieke, G. H., Rieke, M. J., McCarthy, D. W., Haemmerle, V., Haller, J., McLeod, B., Lebofsky, L. A., Marcialis, R., Holberg, J. B., Landau, R., Carrasco, L., Elias, J., Buie, M. W., Dunham, E. W., Persson, S. E., Boroson, T., West, S., French, R. G., Harrington, J., Elliot, J. L., Forrest, W. J., Pipher, J. L., Stover, R. J., Brahic, A., Grenier, I., 1997. Structure of SaturnÕs Mesosphere from the 28 Sgr Occultations. Icarus, 130, 404.

Hubbard, W. B., 1999, NOTE: Gravitational Signature of Jupiter's Deep Zonal Flows. Icarus, 137, 357.  

Ingersoll, A. P., 1970. Motions in Planetary Atmospheres and the Interpretation of Radio Occultation Data. Icarus, 13, 34. 

Ingersoll, A. P. and Pollard, D., 1982, Motion in the interiors and atmospheres of Jupiter and Saturn - Scale analysis, anelastic equations, barotropic stability criterion. Icarus, 52, 62. 

Jacobson, R. A. 2003, JUP230 orbit solution, http://ssd.jpl.NAsa.gov/?gravity\_fields\_op 

Jacobson, R. A., Antreasian, P. G., Bordi, J. J., Criddle, K. E., Ionasescu, R., Jones, J. B., Mackenzie, R. A., Meek, M. C., Parcher, D., Pelletier, F. J., Owen, Jr., W. M., Roth, D. C., Roundhill, I. M. and Stauch,
J. R.,  2006. The Gravity Field of the Saturnian System from Satellite Observations and Spacecraft Tracking Data. ApJ, 132, 2520.

Kaspi, Y., Hubbard, W. B., Showman, A. P. and Flierl, G. R., 2010. Gravitational signature of Jupiter's internal dynamics. Geophys. Res. Lett., 37, L01,204. 

Kaspi, Y. 2013. Inferring the depth of the zonal jets on Jupiter and Saturn from odd gravity harmonics. Geophys. Res. Lett., in press. 

Kong, D., Zhang, K., Schubert, G., 2012. On the variation of zonal gravity coefficients of a giant planet caused by its deep zonal flows. ApJ, 748, 143. 

Kramm, U., Nettelmann, N., Redmer, R. and Stevenson, D. J., 2011. A\&A, 528, 7.

Leconte, J. and Chabrier, G., 2012. A new vision on giant planet interiors: the impact of double diffusive convection. Astronomy and Astrophysics, 540, 20.

Lindal, G. F., Wood, G. E. , Levy, G. S., Anderson, J. D. , Sweetnam, D. N., On the degeneracy of the tidal Love number k$_2$ in multi-layer planetary models: application to Saturn and GJ 436b. A\&A, 528, 7.

Lindal G. F., Wood G. E., Levy G. S., Anderson J. D., Sweetnam D. N., et al. 1981. The atmosphere of Jupiter - an analysis of the Voyager radio occultation measurements. J. Geophys. Res. 86, 8721.

Lindal, G. F., Sweetnam, D. N., and Eshleman, V. R., 1985. The atmosphere of Saturn - an analysis of the Voyager radio occultation measurements. ApJ, 90, 1136. 

Lindal, G. F. 1992, The atmosphere of Neptune - an analysis of radio occultation data acquired with Voyager 2. ApJ, 103, 967. 

Read, P. L., Dowling, T. E. and Schubert, G., 2009. Rotation periods of Jupiter and Saturn from their atmospheric planetary-wave configurations, 2009. Nature, 7255, 608. 

Riddle, A. C. and Warwick, J. W. 1976, Redefinition of System III longitude, Icarus, 27, 457. 

Sanchez-Lavega, A., Rojas, J. F., Sada, P. V., 2000. Saturn's Zonal Winds at Cloud Level. Icarus, 147, 405. 

Saumon, D. and Guillot,T., 2004. Shock Compression of Deuterium and the Interiors of Jupiter and Saturn. ApJ,  609, 1170. 

Sterenborg, M. G. and Bloxham, J., 2010. Can Cassini magnetic field measurements be used to find the rotation period of Saturn's interior? GRL, 37, L11201. 

Zharkov, V. N. and Trubitsyn, V. P. 1978. Physics of planetary interiors (Astronomy and Astrophysics Series, Tucson: Pachart, 1978). 


\newpage

\begin{figure}
    \centering
    \includegraphics[width=4.8in]{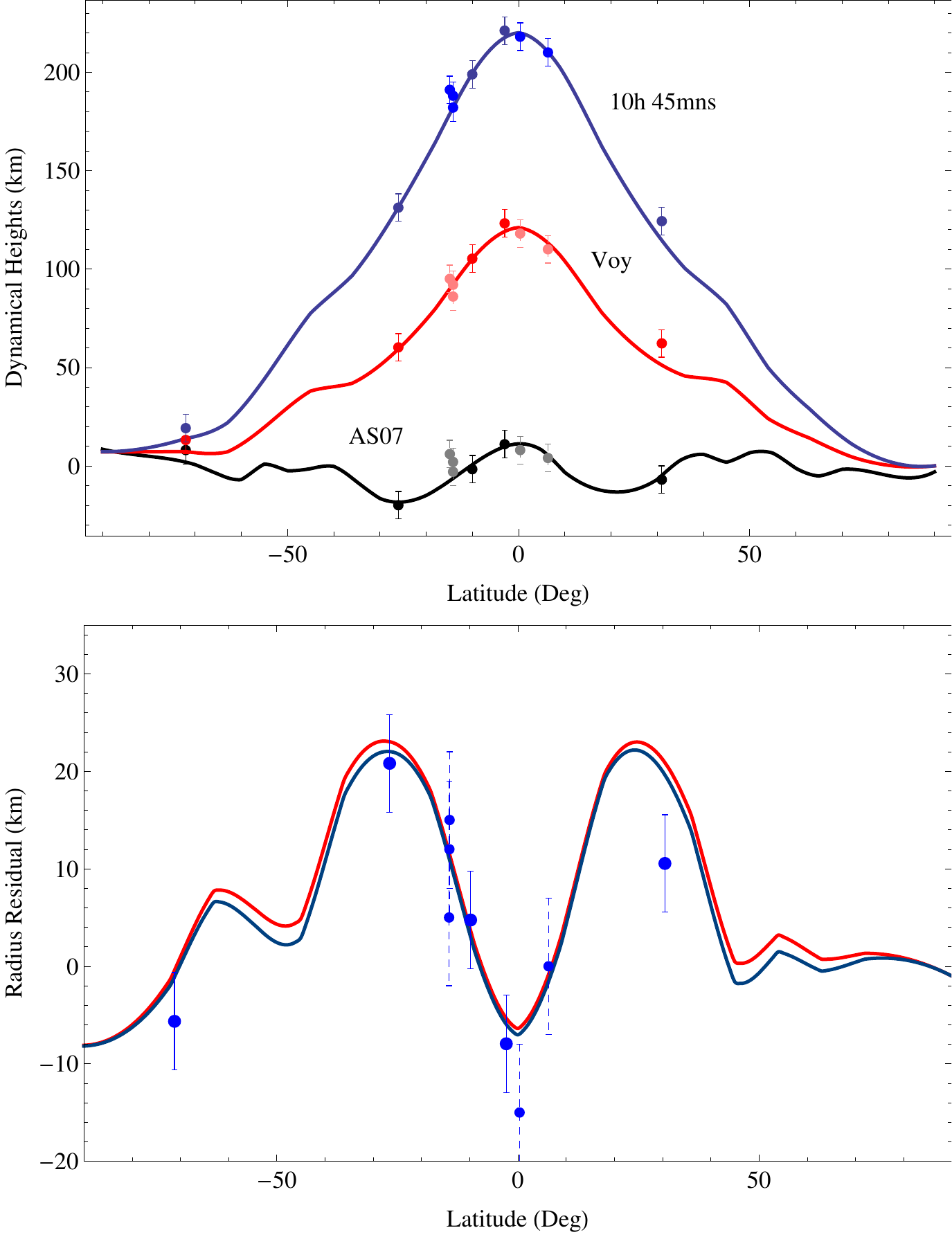}
    \caption[sat]{Top: Saturn's altitudes of the 100 mbar isobaric surface above a reference geoid (dynamical heights, $h(\theta)$) vs. altitude for rotation periods of 10h 32mns 35s (AS07, black), 10h 39mns 22.4s (Voyager, red), and 10h 45mns 24s (blue). The dark-color circles represent radii obtained by radio occultation measurements with the Pioneer 11, Voyager 1, and Voyager 2 spacecrafts (Lindal et al., 1985)  with an error of 7 km. The light-color circles are the radii obtained from stellar occultation measurements (Hubbard et al., 1997).
    Bottom: Radius residuals as a function of latitude when the reference (i.e., Saturn's physical shape) is defined by the shape of a geoid with AS07 rotation period. The red and blue colors correspond to rotation periods of 10h 39mns 22.4s and 10h 45mns 24s, respectively. The light blue circles are the radio (solid) and stellar (dashed) occultation residuals.}
\end{figure}

\begin{figure}
    \centering
    \includegraphics[width=2.2in]{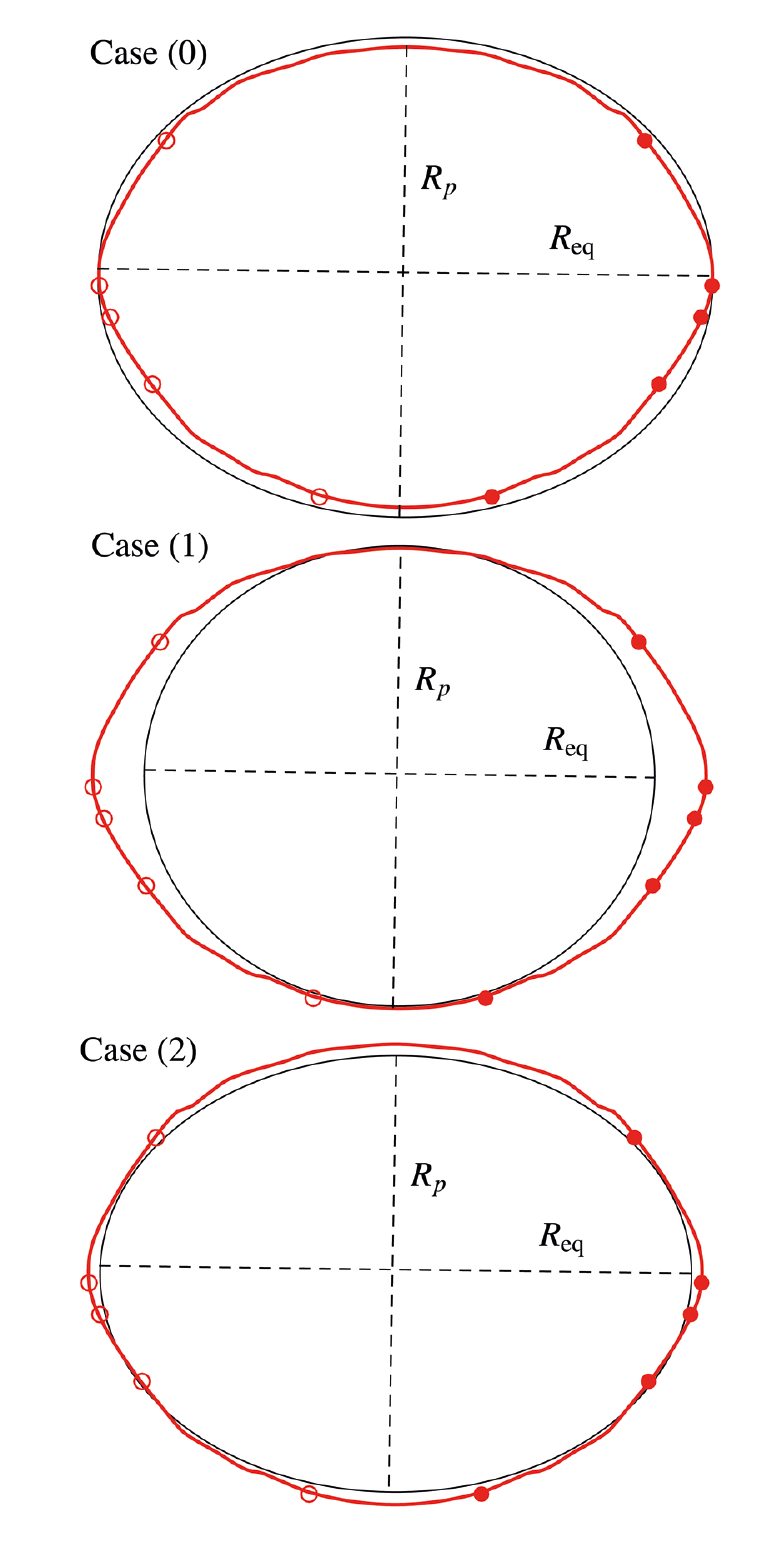}
    \caption[sat]{Sketches of the three cases we consider. Case (0) Saturn's equatorial radius is fixed to the observed value. Case (1): Saturn's polar radius is fixed to its estimated value; the equatorial radius is then found to reproduce Saturn's mean volume as determined from the five radio occultation measurements. Case (2): both the equatorial and polar radii of Saturn are varied to reproduce its measured volume (mean radius) from occultation measurements. Again, the circles represent the radii obtained by radio occultation measurements. Note that the sketches are not to scale - the effect of the winds has been exaggerated by a factor 100 in order to be noticeable.}
\end{figure}

\begin{figure}
    \centering
    \includegraphics[width=5.in]{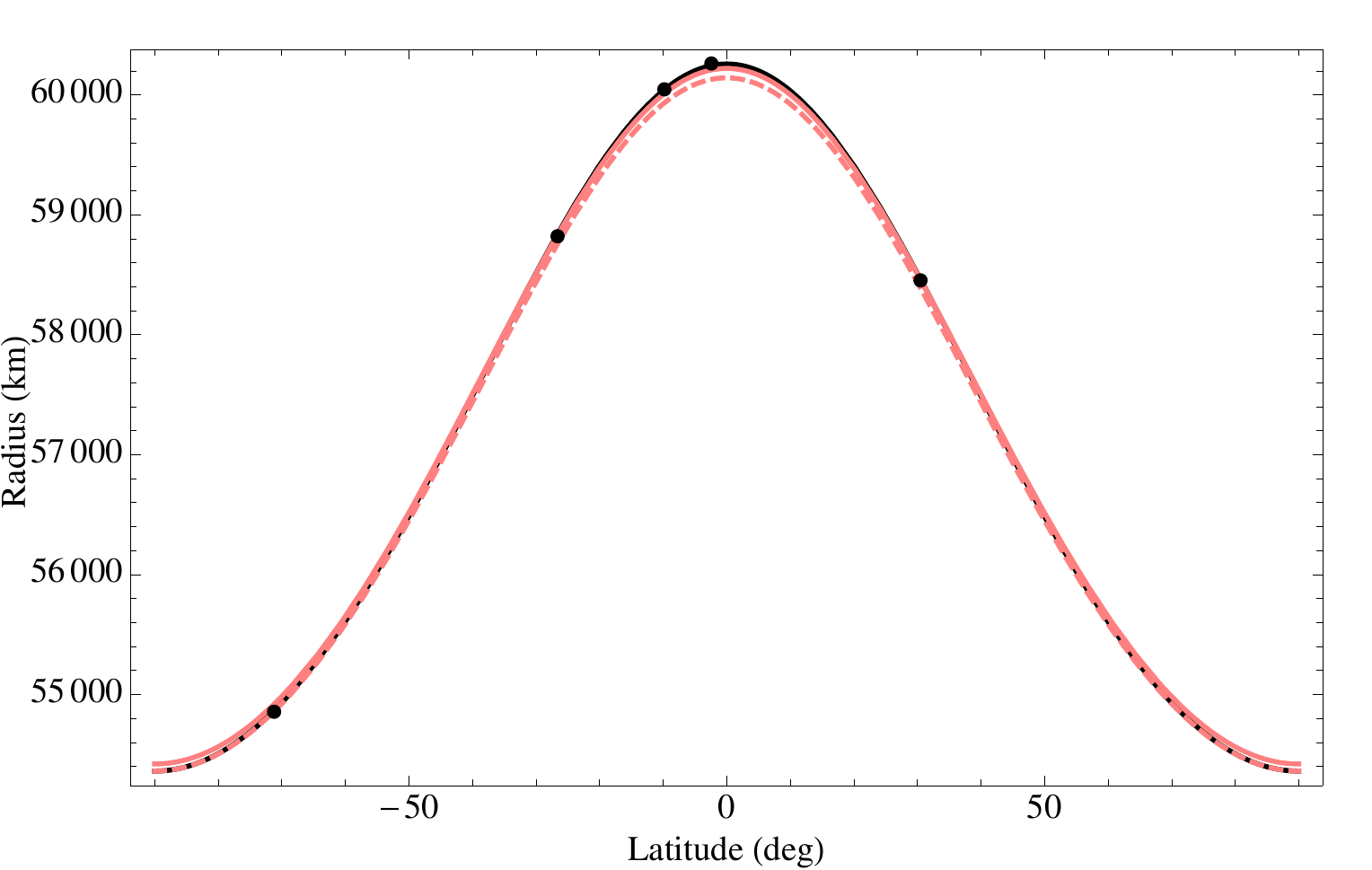}
    \caption[sat]{The shape of a reference geoid with AS07 solid-body rotation period (black). The black circles are the five radio occultation measurements from the Pioneer 11, Voyager 1, and Voyager 2 spacecrafts (Lindal et al., 1985). The red curves correspond to reference geoids with Voyager's rotation period of 10h 39mns 22.4s; the dashed and solid red lines correspond to Case (1) and Case (2), respectively.}
\end{figure}

\begin{figure}
    \centering
    \includegraphics[width=4.in]{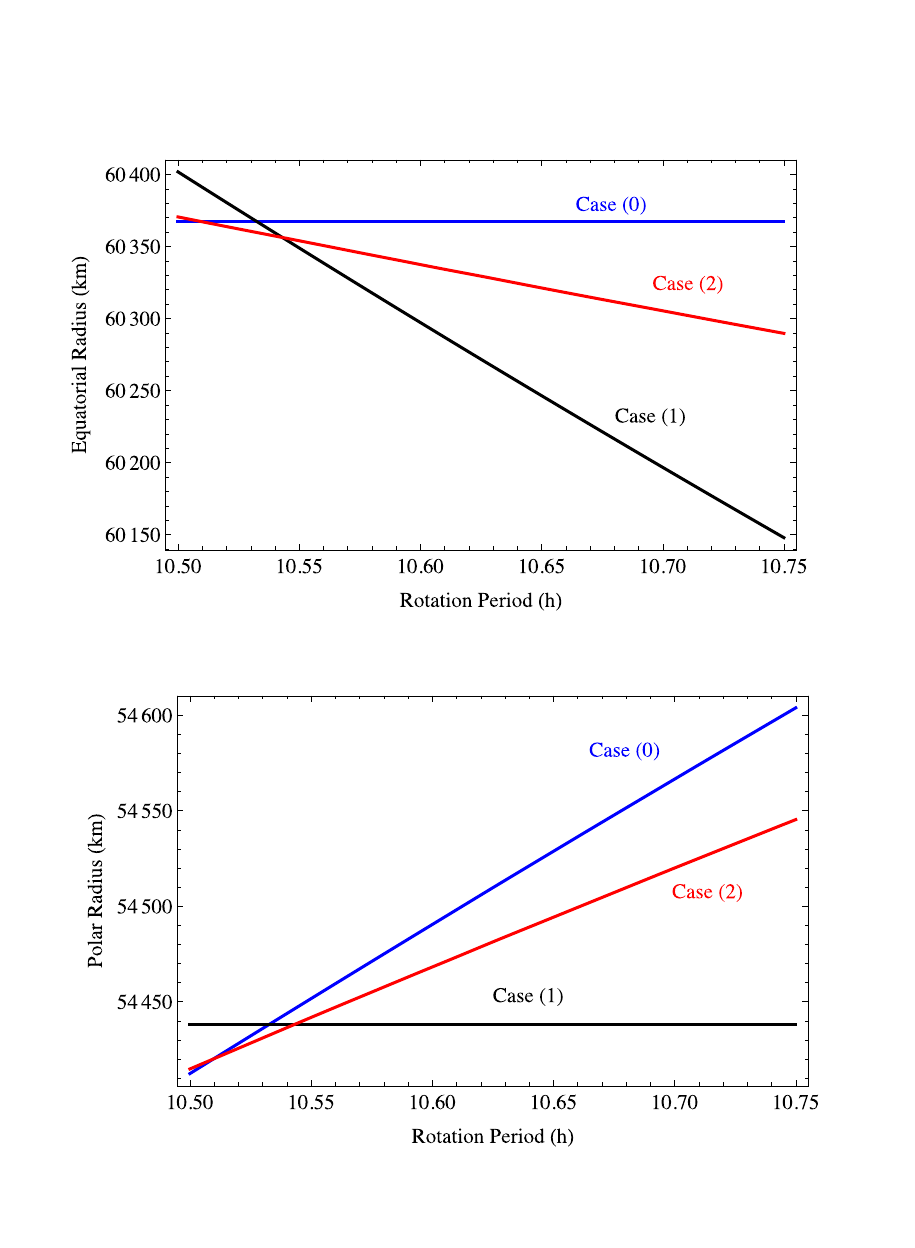}
    \caption[sat]{Saturn's equatorial (top) and polar (bottom) radii as a function of rotation period. The blue, black and red curves correspond to Case (0), Case (1) and Case (2), respectively. The calculations correspond to the 100 mbar pressure-level. It is suggested that the differences between the Case (1) and Case (2) should be taken as the uncertainty in Saturn's equatorial/polar radius (see text).}
\end{figure}

\begin{figure}
    \centering
    \includegraphics[width=5.5in]{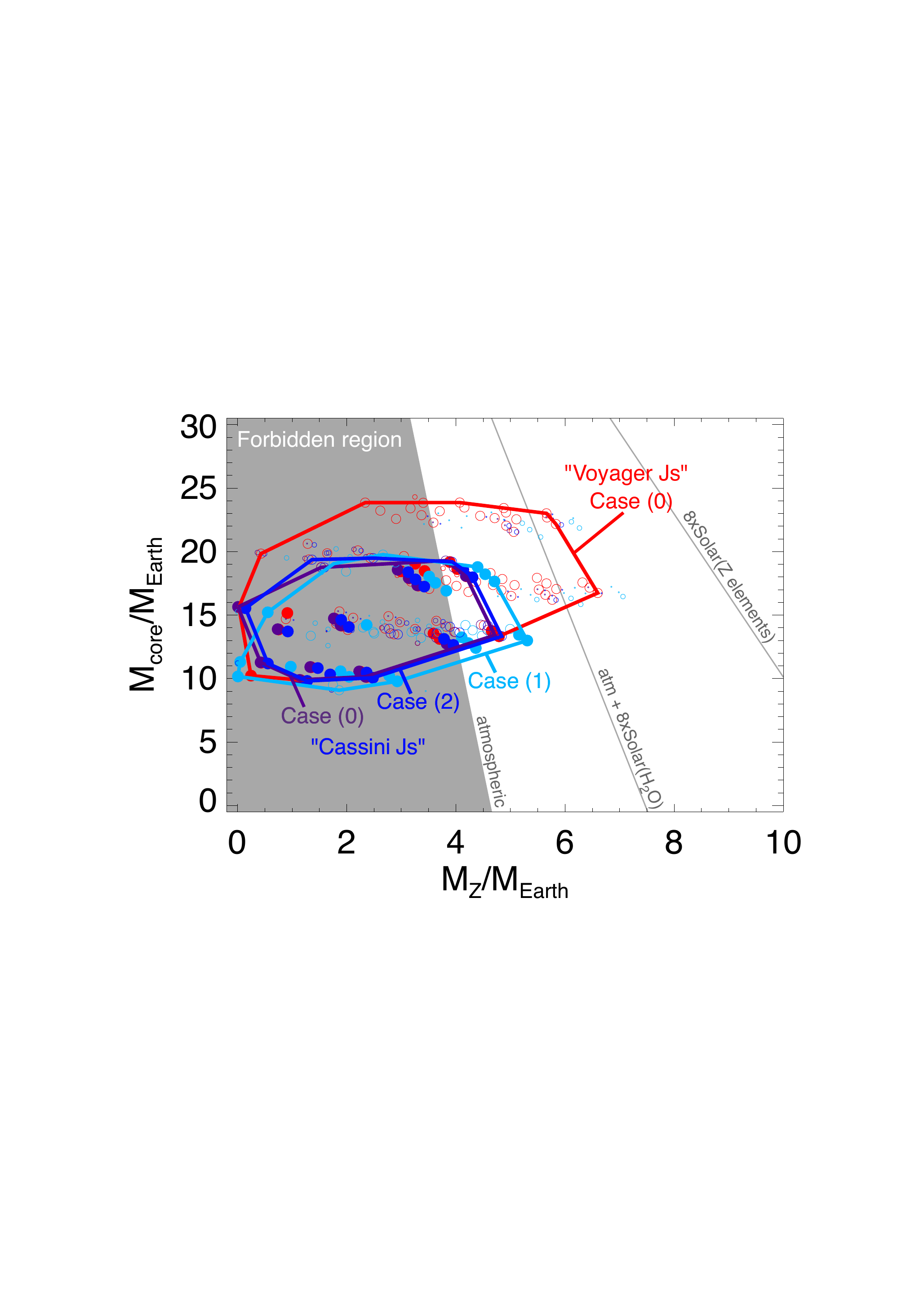}
    \caption[sat]{The derived values of Saturn's core mass (M$_{\text{core}}$) are shown as a function of the mass of heavy elements in the envelope (M$_{\text Z}$) for interior models matching the available observational constraints. Solutions for $P_{{\text{transition}}}$ = 1 Mbar using the Voyager rotation period with Voyager's $J$s, and model Case (0) (red), and for Cassini $J$s and models Case (0) (purple), Case (1) (blue), Case (2) (light blue). The grey region corresponds to available contraints from atmospheric abundances in CH$_4$, NH$_3$ and H$_2$S which represent a minimum value when the envelope is assumed homogeneous in heavy elements. The first grey line represent that value to which an abundance of H$_2$O equals to 8 times the solar value (Asplund et al. 2009) has been added. The second grey line assumes that all heavy elements (including rock-bearing elements) have an 8 times solar enrichment.  }
\end{figure}

\clearpage
\begin{figure}
    \centering
    \includegraphics[width=5.5in]{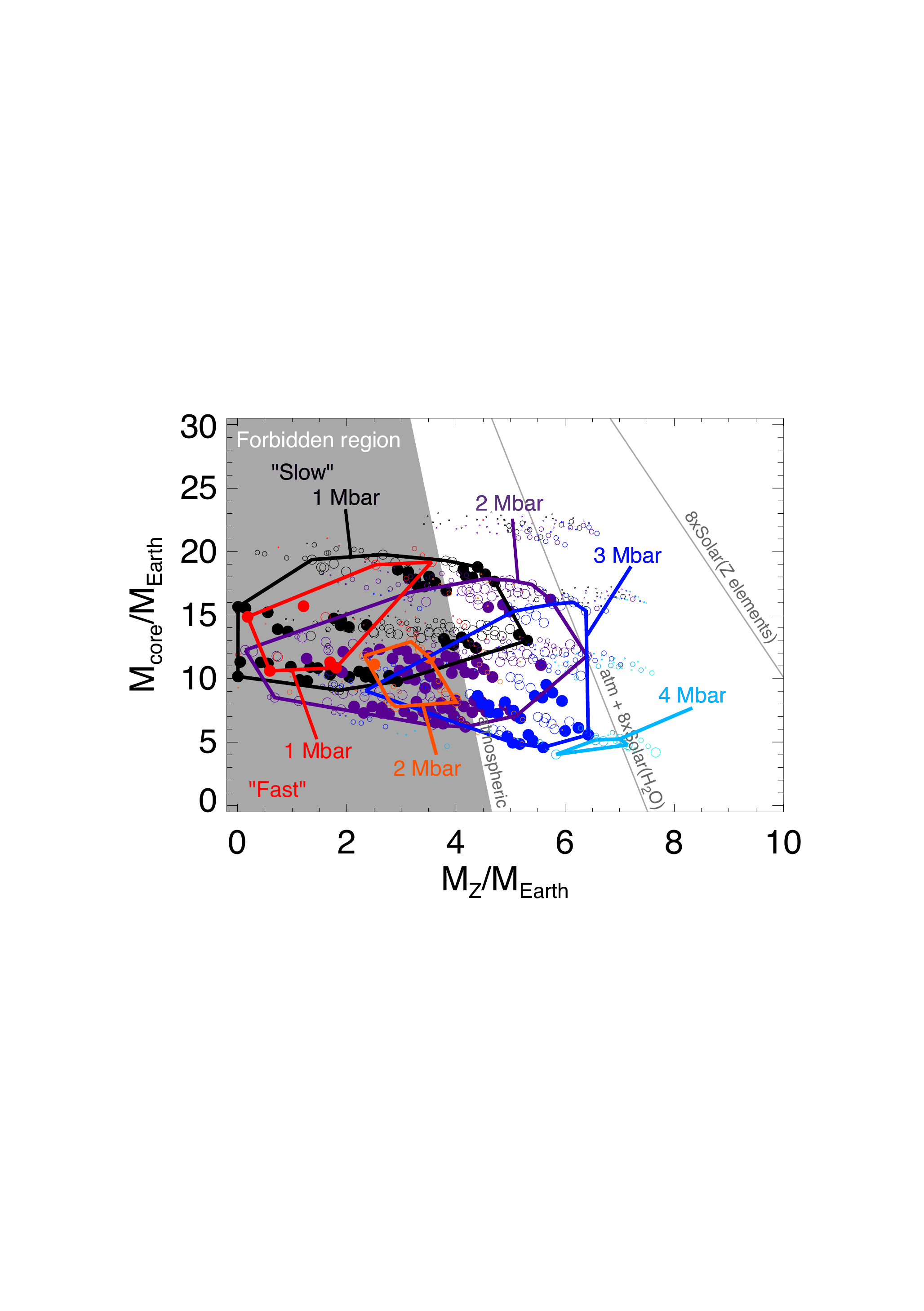}
    \caption[sat]{Same as figure 5 but when using the Cassini $J$s, combining Case (0), Case (1), and Case (2): (i) Voyager rotation period (``Slow'') and $P_{{\text{transition}}}$ = 1 Mbar (black), 2 Mbar (purple), 3 Mbar (blue), 4 Mbar (light blue). 
(ii) AS07 rotation period (``Fast'') and $P_{{\text{transition}}}$ = 1 Mbar (red), 2 Mbar (orange). No solutions can be found for AS07 rotation period with higher $P_{{\text{transition}}}$.}
\end{figure}

\end{document}